\documentclass[letter]{aa} 

\usepackage{graphicx}    
\usepackage{multirow}
\usepackage{txfonts}
\newcommand{\OI}{O\,{\sc i}}
\newcommand{\CI}{C\,{\sc i}}
\newcommand{\CII}{C\,{\sc ii}}

\begin{document}

\title{High-velocity hot CO emission close to Sgr\,A$^*$}
           \subtitle{\textit{Herschel}/HIFI\thanks{\textit{Herschel} is an ESA space observatory with science instruments provided by European-led Principal Investigator consortia  with important participation from NASA.}$^,$\thanks{Includes IRAM\,30\,m telescope
    observations. IRAM is supported by INSU/CNRS (France), MPG (Germany), and IGN (Spain).} submillimeter spectral survey toward Sgr\,A$^*$}

 \titlerunning{High-velocity mid-$J$ CO lines toward Sgr\,A$^*$} 
\authorrunning{Goicoechea, Santa-Maria, Teyssier et al.}

    \author{J. R. Goicoechea\inst{1}
    \and
    M. G. Santa-Maria\inst{1}
          \and
          D. Teyssier\inst{2}
          \and
          J. Cernicharo\inst{1}
          \and 
          M. Gerin\inst{3}
                    \and
          J. Pety\inst{3,4}
          }

   \institute{Instituto de F\'{\i}sica Fundamental (CSIC). Calle Serrano 121, 28006, Madrid, Spain.
   \email{javier.r.goicoechea@csic.es}
         \and
Telespazio Vega UK Ltd for ESA/ESAC. Urbanizaci\'on
Villafranca del Castillo, Villanueva de la
Ca\~{n}ada, E-28692 Madrid, Spain.         
                  \and
Sorbonne Universit\'e, Observatoire de Paris, Universit\'e PSL, \'Ecole Normale Sup\'erieure, CNRS, LERMA, F-75014, Paris, France.
\and
Institut de Radioastronomie Millim\'etrique (IRAM), 300 rue de la Piscine, F-38406 Saint Martin d’H\`eres, France.
 }

   \date{Received June 20, 2018; accepted July 18, 2018}

   
  \abstract{ 
   The properties of molecular gas, the fuel that forms stars,  inside the cavity of the circumnuclear~disk (CND) are not well constrained.   
We present results of a velocity-resolved submillimeter   scan  
(\mbox{$\sim$480 to 1250~GHz}) and [\CII]\,158\,$\mu$m line observations
 carried out with \textit{Herschel}/HIFI toward Sgr~A$^*$; these results are 
   complemented  by a $\sim$2$'$$\times$2$'$  \mbox{$^{12}$CO ($J$=3-2)} map  taken with the \mbox{IRAM\,30 m telescope} at $\sim$7$''$ resolution. 
We report the presence of \mbox{high positive-velocity} emission 
(up to about \mbox{$+$300 km\,s$^{-1}$})   
   detected in the wings of \mbox{$^{12}$CO $J$=5-4} to 10-9 lines. This wing component is also seen in \mbox{H$_2$O~(1$_{1,0}$-1$_{0,1}$), } a tracer of hot molecular gas; in [\CII]158\,$\mu$m, an~unambiguous tracer of \mbox{UV radiation}; but not in  [\CI]\,492,\,806\,GHz. 
This first measurement of the \mbox{high-velocity} $^{12}$CO \mbox{rotational ladder} toward Sgr~A$^*$ adds more evidence that hot molecular gas exists inside the cavity of the CND,  relatively close to the supermassive black hole ($<$~1\,pc). Observed by ALMA, this velocity range appears as a collection of \mbox{$^{12}$CO ($J$=3-2)} cloudlets
 lying in a very  harsh environment that is pervaded by intense  UV radiation fields, shocks, and affected by strong  gravitational shears. We constrain the physical conditions of the  high positive-velocity 
CO gas component by comparing with \mbox{non-LTE} excitation and radiative transfer models. We infer $T_{\rm k}$$\simeq$400~K to 2000~K for \mbox{$n_{\rm H}$$\simeq$(0.2-1.0)$\cdot$10$^5$~cm$^{-3}$}.
These results point toward the important role of stellar UV radiation, but
we show that radiative heating alone cannot explain the excitation  of this 
\mbox{$\sim$10-60\,$M_{\odot}$} component of hot molecular gas  inside the central cavity.
Instead, strongly irradiated shocks are  promising candidates.}

   \keywords{ISM: clouds --- ISM: kinematics and dynamics --- ISM: molecules --- Galaxy: center --- radio lines: ISM}

   \maketitle
%

\section{Introduction}

The Galactic center (GC), in particular the neighborhood
of the central supermassive black hole (SMBH) at Sgr\,A$^*$ \citep{Schodel02,Ghez08,Gillessen09}, represents a unique laboratory to understand
the properties of interstellar gas in extreme conditions: intense UV fields and  powerful winds from massive stars, shocks, strong gravitational shears, and complicated orbital
motions 
\citep[][]{Morris96,Genzel10}.
The interstellar material in the innermost region of our Galaxy consists of a  cavity of radius $\sim$1.5~pc containing warm dust and gas, ionized by UV radiation from about 100  \mbox{OB-type} massive  stars of the central cluster \citep[e.g.,][]{Krabbe95}. Some of the ionized gas streamers 
(the \mbox{mini-spiral}) seem to bring material close to the very center \citep[e.g.,][]{Zhao09,Irons12,Lau13}.
Between $\sim$1.5\,pc and $\sim$2.5\,pc, a ring of denser molecular gas and cooler dust  exists 
\citep[the CND, e.g.,][]{Gusten87,Bradford05,Montero09,Requena12,Mills13}.

Owing to hostile  conditions inside the cavity of the CND, the presence of molecular
gas was originally not expected. 
Recent studies, however, point toward  its existence   \citep[e.g.,][]{Herrnstein02,Goicoechea13,Ciurlo16,Moser17,Mills17,Yusef17}.
Indeed, obtaining high-resolution spectral images of the molecular gas emission close to
Sgr\,A$^*$ is currently feasible with radio interferometers such as ALMA \citep[e.g.,][]{Moser17,Yusef17}. In particular, \citet{Goicoechea18} have detected a collection of \mbox{$^{12}$CO ($J$=3-2)} cloudlets  emitting at higher positive velocities than reported before (up to \mbox{v$_{\rm LSR}$\,$\simeq$$+$300~\,km$^{-1}$}). Owing to the small cloudlet sizes
($\lesssim$20,000~AU), broad line profiles (\mbox{$\Delta {\rm v}$$\simeq$25-40~km\,s$^{-1}$}), but similar line-velocity centroids over the cavity, these authors have suggested that these cloudlets are leftovers of disrupted clouds that fall into the central 
region. Other dissipative effects such as shocks in colliding streams, cloudlet-cloudlet collisions, colliding stellar winds, or advancing photoionization fronts, perhaps  triggering gas compression and molecular gas formation, may also explain the presence of molecular gas in the cavity  \citep{Goicoechea18}.

Most of the luminosity emitted by molecular gas heated by shocks or illuminated by strong UV fields  (leading to $T_{\rm k}$$>$100\,K)  is radiated in
\mbox{mid-$J$} and \mbox{high-$J$} lines at  submillimeter and far-IR wavelengths  that are inaccessible to ground-based telescopes. In particular, observations of 
the CO rotational ladder \mbox{\citep[e.g.,][]{Indriolo17,Joblin18}}  are critical
tools to determine the heating mechanisms and nature of the  molecular gas in the GC and, in a broader context, to better understand the 
emission from distant  active and star-forming extragalactic nuclei \citep[e.g.,][]{Werf10,Rangwala11}.

\section{Observations and data reduction}

The spectral scan toward Sgr~A$^*$ 
 was carried out with the heterodyne receiver HIFI  \citep{Graauw10} on board \mbox{\textit{Herschel}} \citep{Pilbratt10}. 
Owing to intense molecular line emission around the CND area, these observations  were performed in the \mbox{\textit{Spectral Scan Load Chop}} mode using an internal cold load as reference.
We used the Wide Band Acousto-Optical Spectrometer, which provides a spectral resolution of
 1.1~MHz and instantaneous bandwidths of 4 (2.4)~GHz in bands 1-5 \mbox{(6-7)}. 
The total observing time was $\sim$40~h. 
Some of the frequency ranges were affected by isolated spurious signals, together with distorted residual baselines due to imperfect gain drift correction. While these signals were treated by the HIFI pipeline through dedicated line masking, the second artifact was corrected by fitting polynomial baselines
on each  spectrum prior to deconvolution \citep[][]{Comito02}. Data calibration was performed in HIPE \citep{Ott10}. Deconvolution allows the recovery of both line and continuum on a single side-band (SSB) intensity and frequency scale, allowing accurate absorption and emission  studies. 
We adopted beam efficiencies and half power beam widths (HPBW) from the latest HIFI beam 
model\footnote{https://www.cosmos.esa.int/web/herschel/legacy-documentation-hifi} 
and used the main-beam temperature scale ($T_{\rm mb}$).
The rms of the SSB spectra (per 2~km\,s$^{-1}$ channel) are $\sim$15, $\sim$25, $\sim$40, $\sim$60, and $\sim$120\,mK for bands~1 to 5, respectively, and $\sim$175\,mK for the [\CII]158\,$\mu$m line. 
Table~\ref{table:detected_line}  lists the total integrated line intensities 
with global uncertainties up to $\simeq$15$\%$, the  mean velocity
($<$v$_{\rm LSR}$$>$), velocity dispersion 
($<$$\Delta$v$>$), and  HPBW at multiple
line frequencies.

\begin{figure}[t]
\centering
\includegraphics[scale=0.1, angle=0]{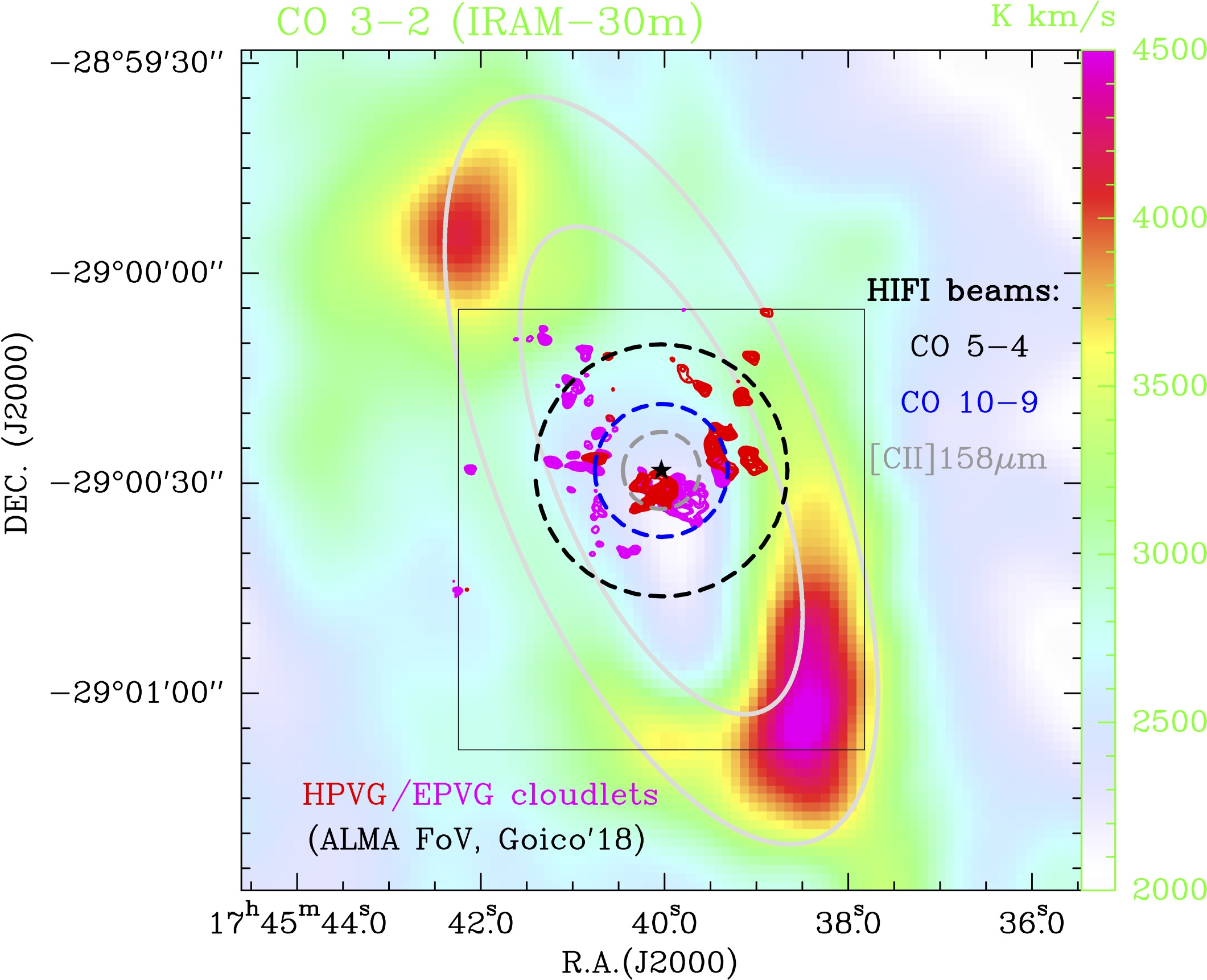} 
\caption{$^{12}$CO~($J$=3-2) integrated intensity map of the inner 2$'$ of the Galaxy obtained with the IRAM\,30 m telescope and dominated by emission from the CND (color scale  in K~km\,s$^{-1}$). The square box shows the field of view observed by ALMA 
in $^{12}$CO~($J$=3-2) revealing  high positive-velocity
 cloudlets inside the cavity \citep{Goicoechea18}. 
The gray curves show representative  orbits 
CND  with 1.5 and \mbox{2.3 pc} de-projected radii.
The black, blue, and gray dashed circles centered at
Sgr~A$^*$ (black star) show HIFI's HPBWs at multiple frequencies
\mbox{(from $\sim$11$''$ to $\sim$40$''$)}.
}
\label{fig:CND-image}  
\end{figure}

We complemented pointed observation by HIFI with a $\sim$2$'$$\times$2$'$ map
of the $^{12}$CO~$J$=3-2 line (345.796~GHz) at  $\sim$7$''$ \mbox{angular resolution}, observed with the IRAM\,30 m telescope (Pico Veleta, Spain)  using the EMIR330 receiver and the FFTS backend at 200~kHz resolution. The map was carried 
by scanning, \mbox{on the fly}, in two orthogonal directions on the sky. \mbox{Owing} to very extended $^{12}$CO emission in the GC, the map was taken using a nearby reference position, at ($+$324$''$,$-$124$''$) with respect to Sgr~A$^*$
at \mbox{(0$''$, 0$''$)}, which we successively calibrated using increasingly distant pointed observations 
($\Delta$R.A.=$+$1000$''$, $+$2000$''$, $+$3000$''$, $+$4000$''$, and $+$5000$''$).
The total integration time was 3.6\,h, leading to a rms of $\sim$1.5~K per
2~km\,s$^{-1}$ channel. The spectra were baselined and gridded to a data cube through convolution with a Gaussian kernel providing a final resolution of $\sim$10$''$.
The total integrated intensity map is shown in Figure~\ref{fig:CND-image}.

\section{Results: Detected lines and wing emission} 

We focus on the brightest detected  emission lines.
These include \mbox{$^{12}$CO ($J$=5 to 10)} rotational lines from  $^{13}$CO,
C$^{18}$O, HCO$^+$, HCN \mbox{($J$=6 to 8)}, CN, and H$_2$O, and 
fine structure lines from ionized (C$^+$) and neutral atomic carbon (C). \mbox{Table~\ref{table:detected_line}} of the Appendix tabulates their spectroscopic and
observational parameters.  Most of these were previously detected at  low spectral resolution with SPIRE and PACS
\citep{Goicoechea13}. The much higher resolution of HIFI allows us to spectrally resolve these complicated \mbox{line profiles} and independently study the various velocity components seen toward the GC. 

In addition to emission, H$_2$O~1$_{1,0}$-1$_{0,1}$
and [\CII]158~$\mu$m lines show absorption features \mbox{below} the continuum level.
 Compared to the CND, the continuum \mbox{flux densities} measured by HIFI toward Sgr~A$^*$ are low, i.e., $\sim$50 and $\sim$100~Jy\,beam$^{-1}$ at 557 and 809~GHz, respectively. This is due to the lower (and hotter) column densities of  dust in the cavity \citep[][]{Etxaluze11}.
The absorption at  v$_{\rm LSR}$$\simeq$0~km\,s$^{-1}$ is produced by local diffuse
clouds of the Galactic  arms \citep[][]{Greaves94} and also by clouds
in the GC itself \citep[e.g.,][]{Goto02}. In addition, the \mbox{H$_2$O~1$_{1,0}$-1$_{0,1}$} line shows absorption at high negative velocities,
v$_{\rm LSR}$$\approx$$-$200 to $-$100~km\,s$^{-1}$
\citep[seen also in ][]{Sandqvist03}, generally attributed to
 \mbox{high negative velocity gas} clouds \citep[$-$180~km\,s$^{-1}$ feature,][]{Gusten81}  and to the near side of the, historically called, expanding molecular ring 
\citep[EMR or \mbox{$-$135~km\,s$^{-1}$ feature},][]{Scoville72}.
These  features are produced by diffuse, low excitation molecular gas in the \mbox{inner} few hundred pc of the Galaxy \citep[][]{Goto02} but they are 
not associated with Sgr~A$^*$.

\begin{figure}[t]
\centering
\includegraphics[scale=0.33, angle=0]{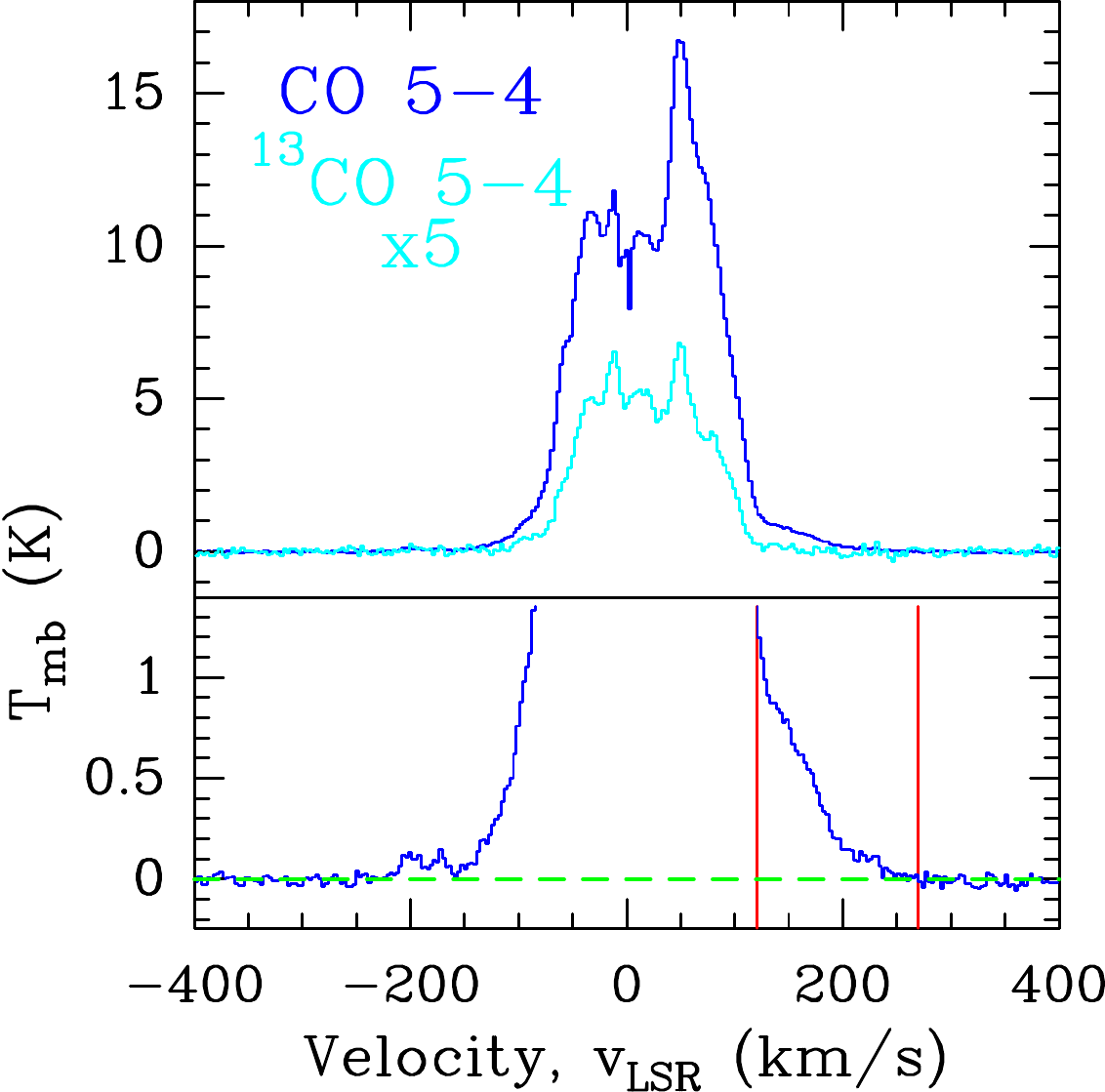} \hspace{0.1cm} 
\includegraphics[scale=0.33, angle=0]{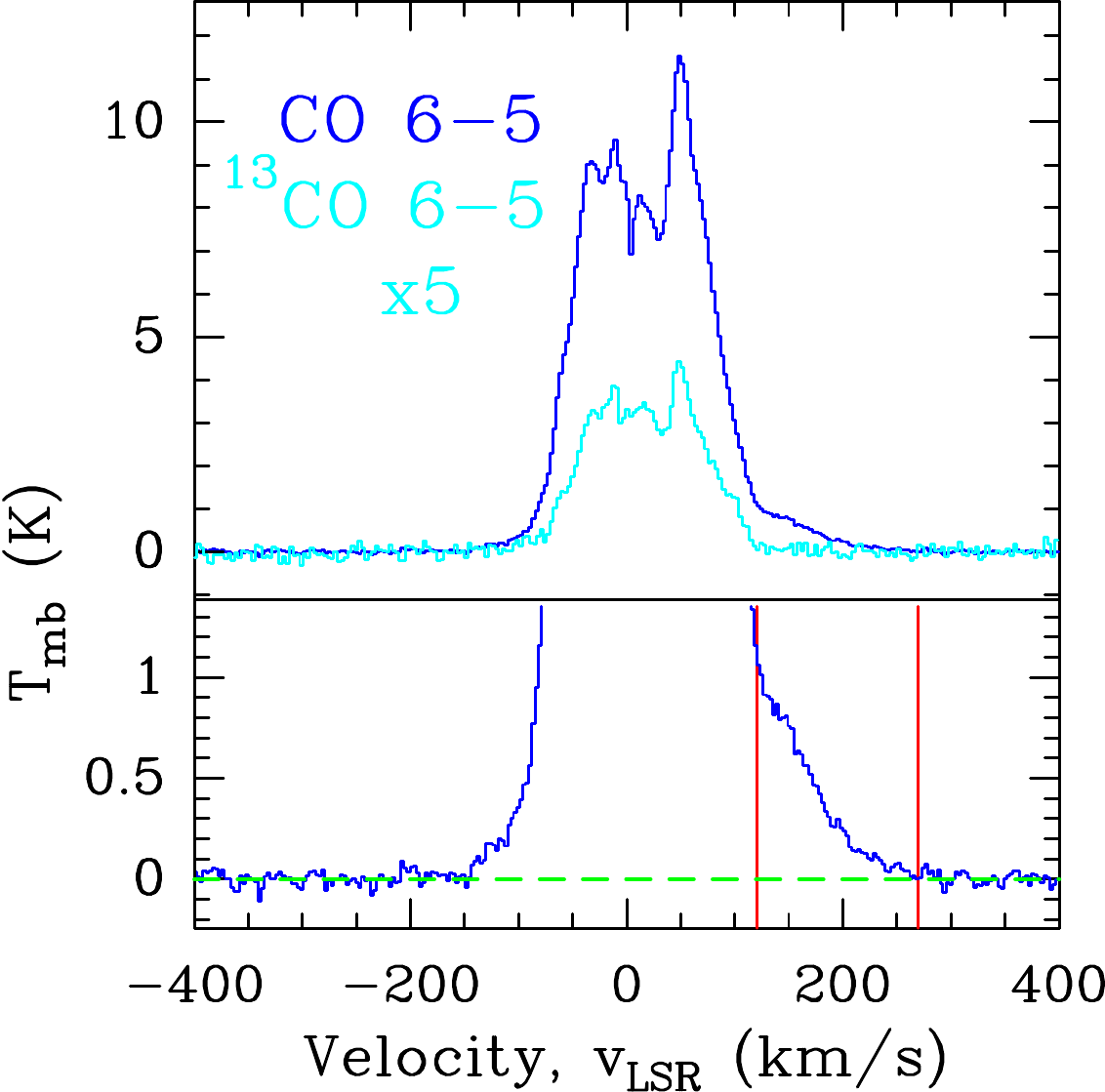} \\
\vspace{0.3cm}
\includegraphics[scale=0.33, angle=0]{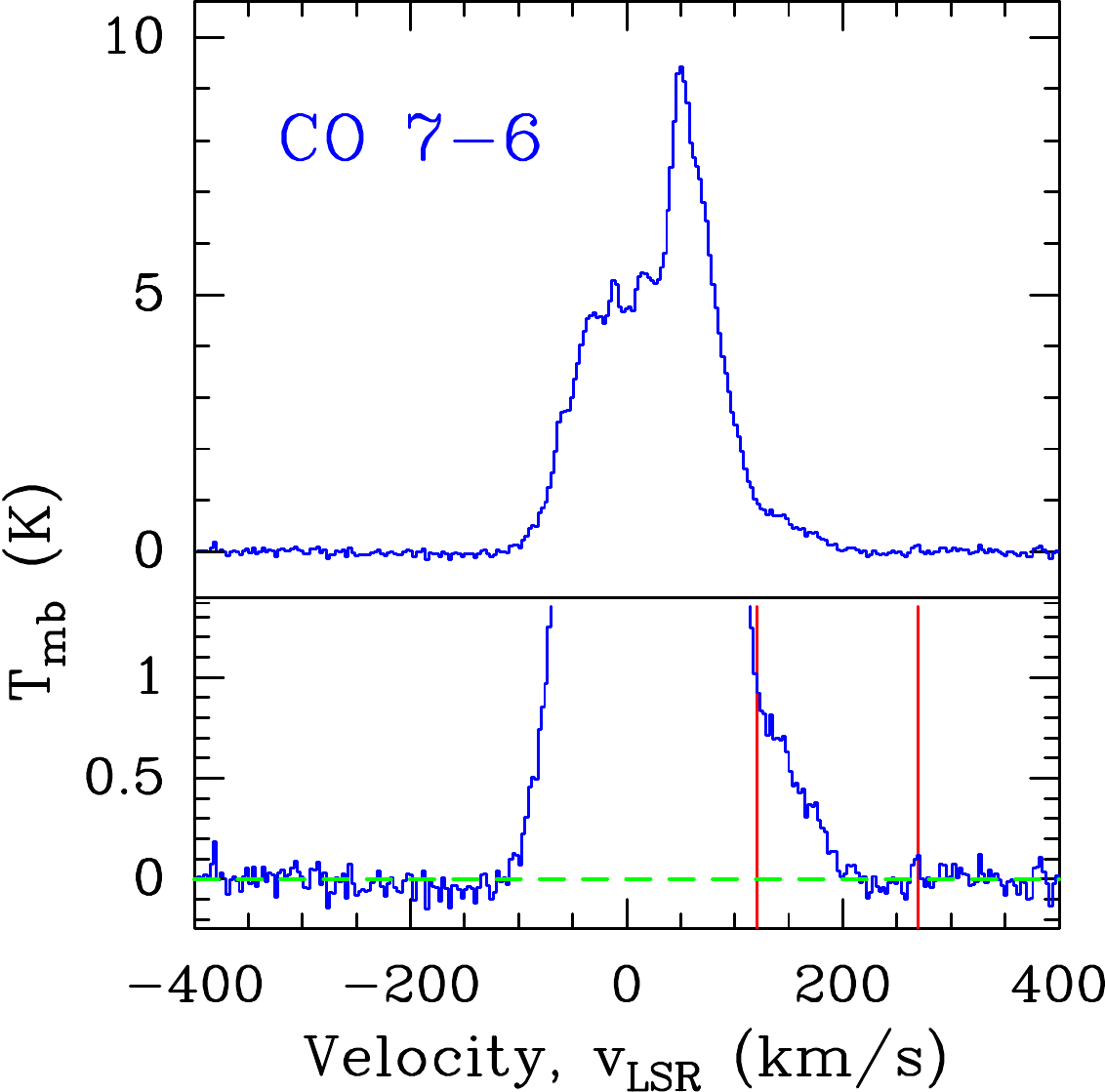} \hspace{0.1cm}
\includegraphics[scale=0.33, angle=0]{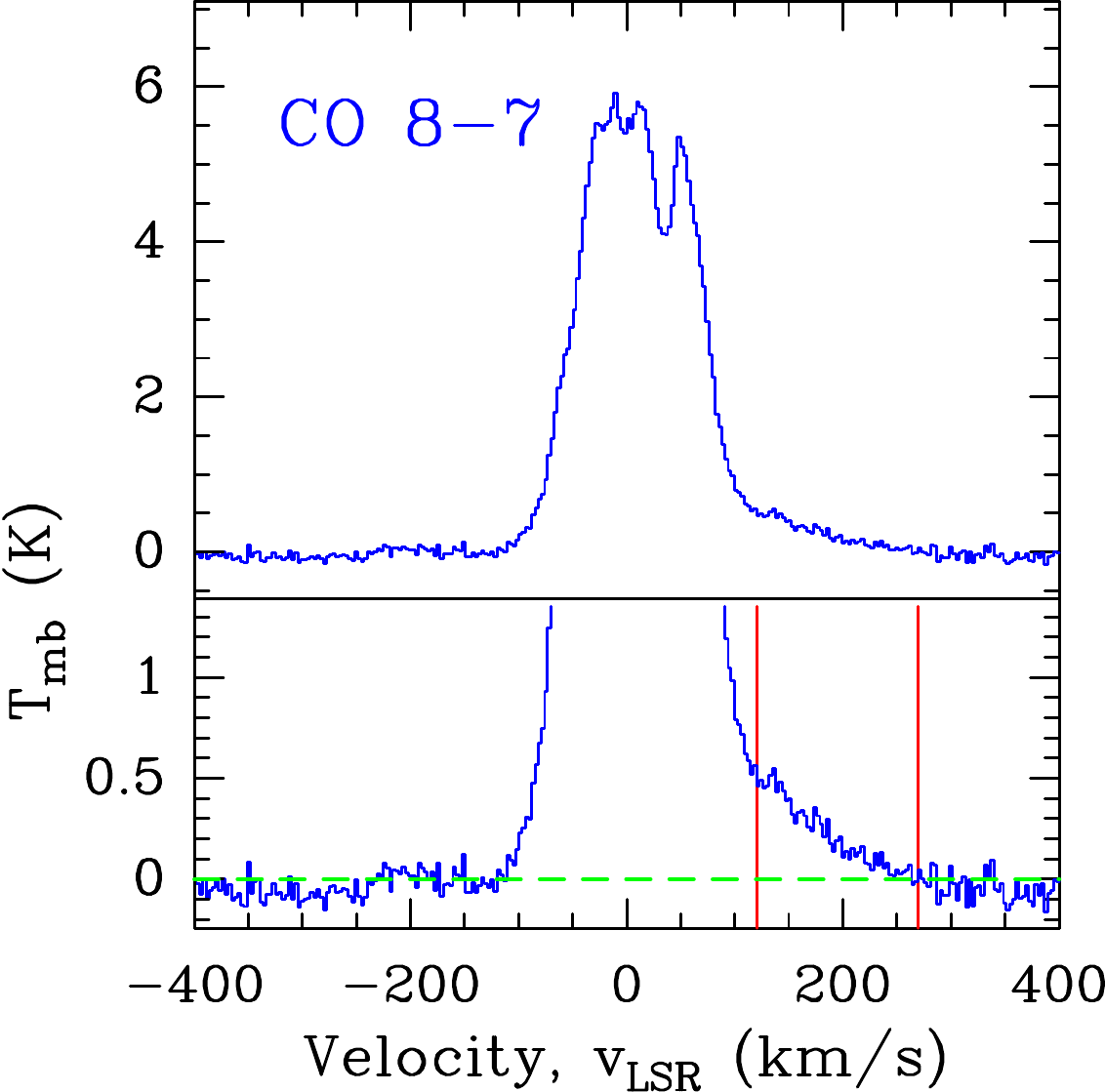} \\
\vspace{0.3cm}
\includegraphics[scale=0.33, angle=0]{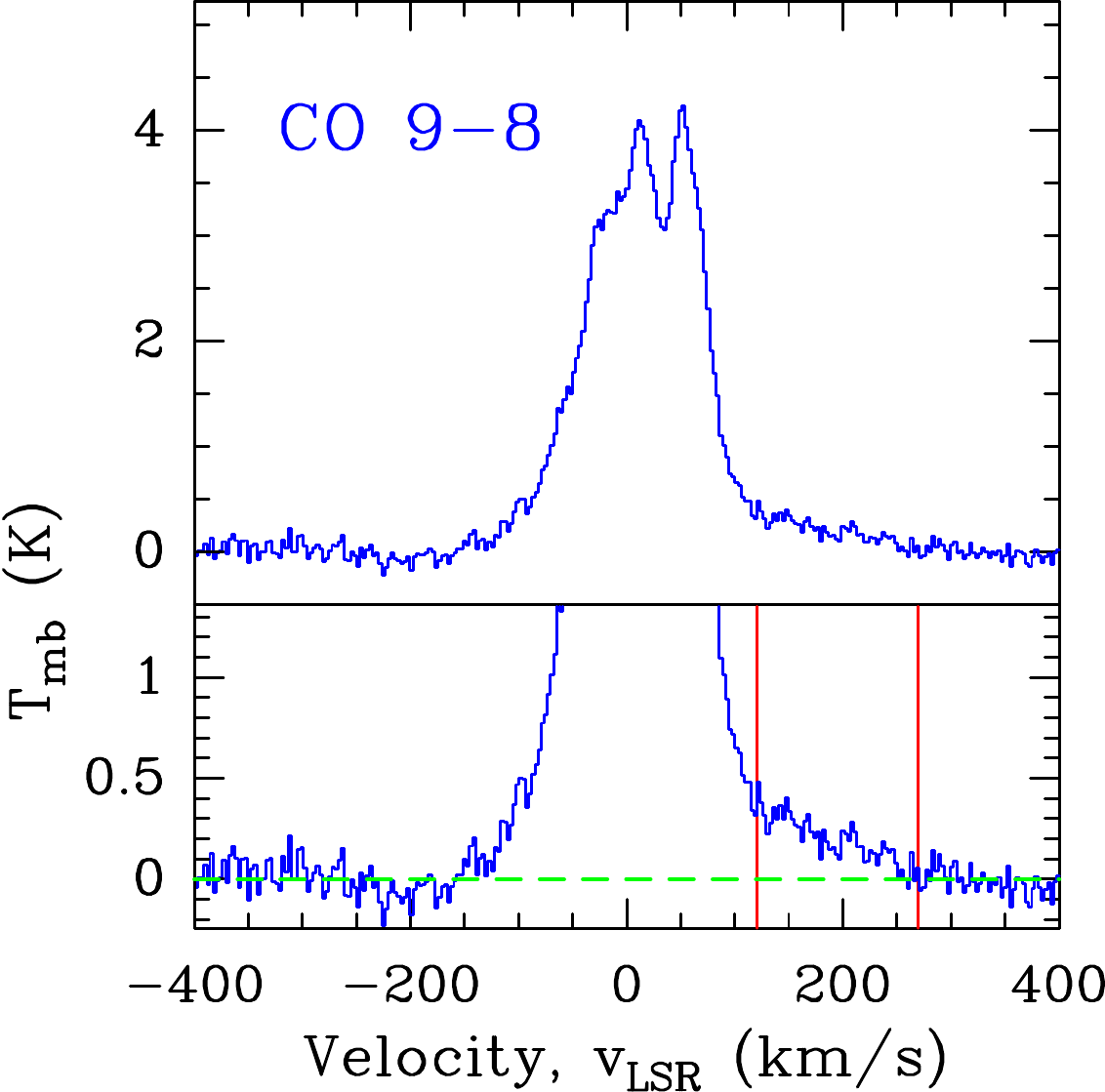} \hspace{0.1cm}
\includegraphics[scale=0.33, angle=0]{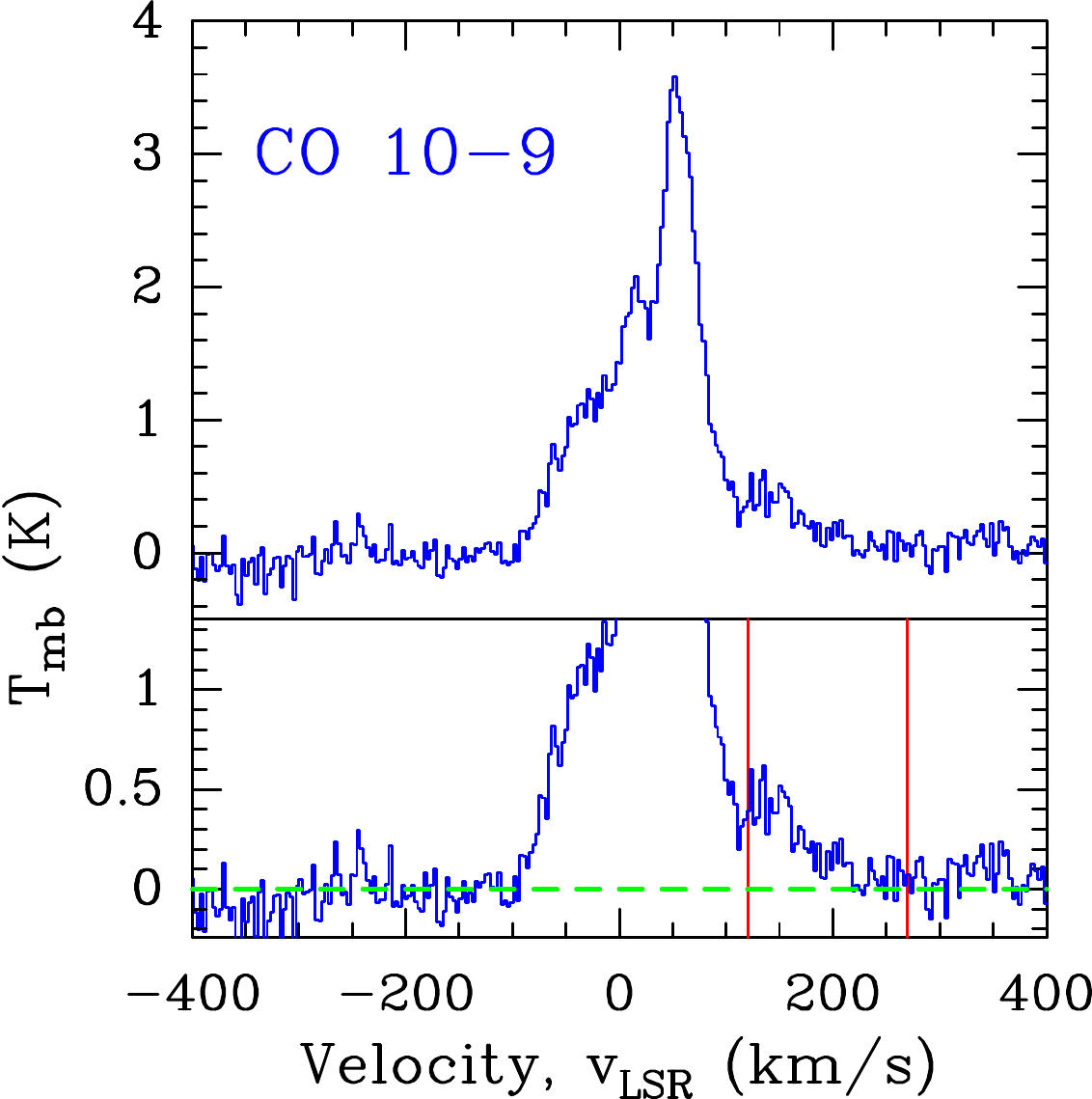} \\
\vspace{0.4cm}
\caption{Mid-$J$ $^{12}$CO and $^{13}$CO lines observed with HIFI toward Sgr\,A$^*$ (baseline subtracted).  
The two red vertical lines at \mbox{v$_{\rm LSR}$\,$\simeq$$+$120 and
$+$270\,km$^{-1}$} delineate the HPVW emission.}
\label{fig:CO-lines}
\end{figure}

In addition to emission at 
\mbox{$|$v$_{\rm LSR}$$|$\,$<$$$150\,km$^{-1}$} (typical  of the CND), 
the mid-$J$ $^{12}$CO  lines display a \mbox{\textit{high positive-velocity wing}}  (HPVW) and have emission up to \mbox{v$_{\rm LSR}$\,$\simeq$$+$270\,km\,s$^{-1}$}
(zooms in Fig.~\ref{fig:CO-lines}). The  HPVW  emission  is faint 
($\sim$500~mK) and is not seen in \mbox{$^{13}$CO ($J$=5-4)}, implying optically thin \mbox{mid-$J$ $^{12}$CO} emission.
This component has not been discussed in previous  observations  of the CND \citep[e.g.,][]{Bradford05,Requena12} but pioneering detection of broad 
\mbox{$^{12}$CO~$J$=7-6}  line widths  
toward  Sgr~A$^*$ suggested the presence of molecular gas within the cavity
\citep{Harris85}. \citet{Goicoechea13} also reported broad 
[\OI]\,63\,$\mu$m emission wings toward  Sgr~A$^*$ (up to $\simeq$300~km\,s$^{-1}$).

Central region ALMA images unveil the presence of small-scale
\mbox{$^{12}$CO ($J$=3-2)} cloudlets inside the   cavity (shown in Fig.~\ref{fig:CND-image}). These cloudlets  emit in the same HPVW  range,
showing independent spectral features at 
\mbox{v$_{\rm LSR}$$=$$+$150-200\,km\,s$^{-1}$} (high positive-velocity gas; HPVG)
and at \mbox{v$_{\rm LSR}$$=$$+$200-300\,km\,s$^{-1}$} (extreme positive velocity gas; EPVG).
From the \mbox{CO~$J$=3-2} maps, we determine an
\mbox{interferometer-alone} to \mbox{single-dish} line flux ratio of $\simeq$0.2 
(HPVW range).
 As we expect the unresolved cloudlets  to be hot 
 \citep[][]{Goicoechea13,Mills17}
 the contrast between emission from the extended background and these cloudlets is likely to increase for the more excited CO lines, leading to larger  flux ratios.
 In the following we attribute the origin
of the \mbox{mid-$J$ $^{12}$CO} HPVW emission  to the same cloudlets. 
Their filling factor in the HIFI beams is also about 0.2.

\begin{figure}[t]
\centering
\includegraphics[scale=0.45]{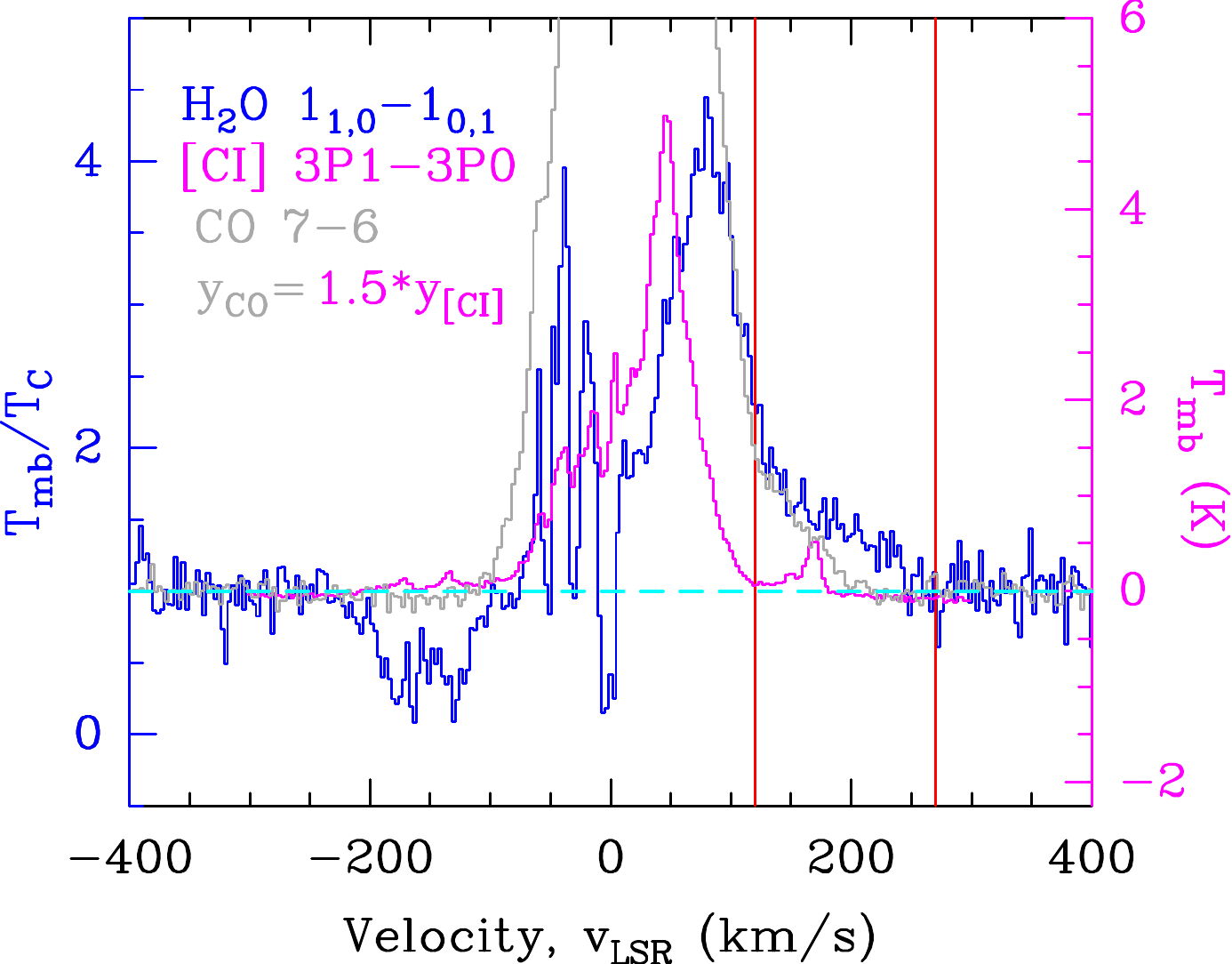}  
\hspace{0.6cm}
\includegraphics[scale=0.45]{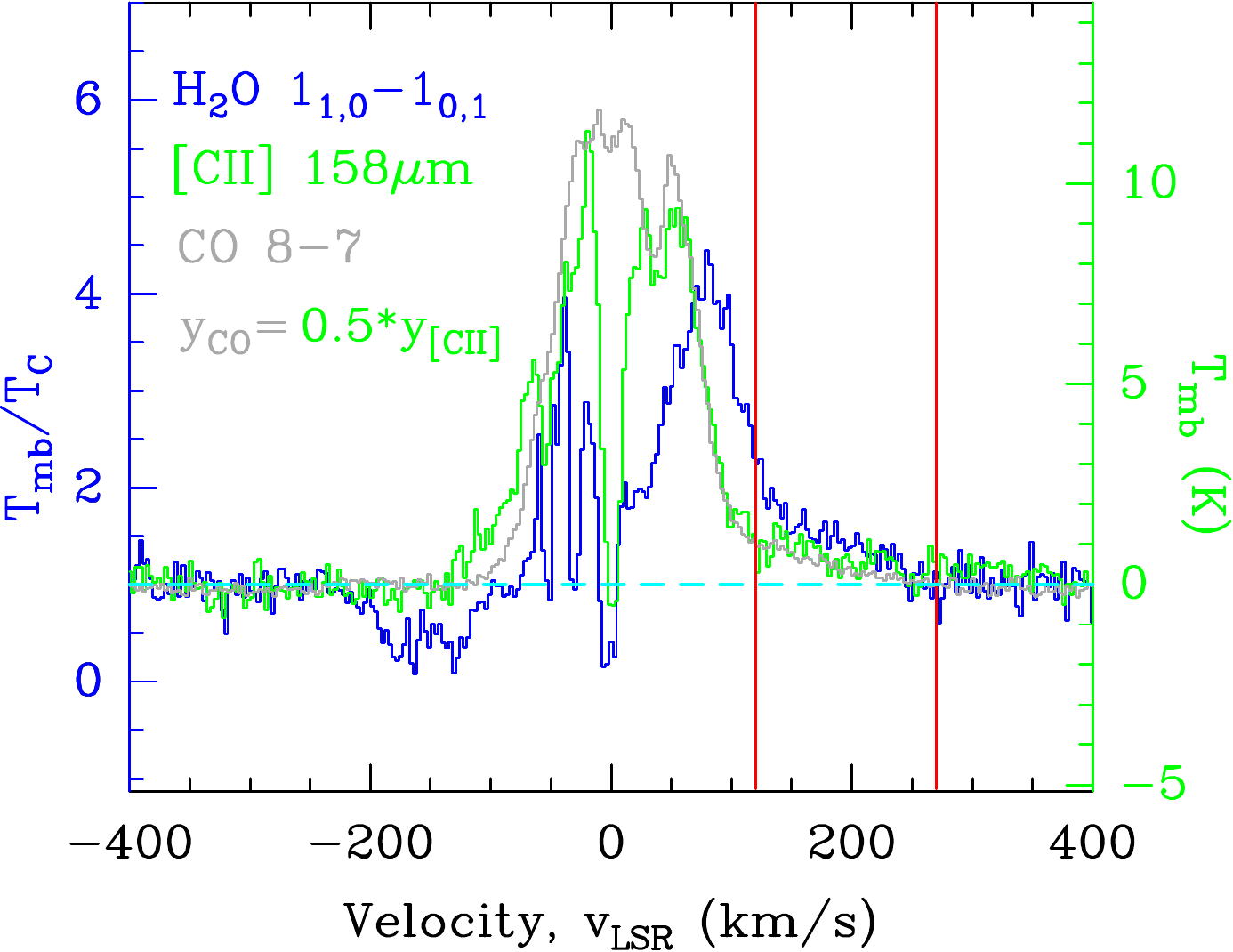} \\
\caption{HIFI multi-line detection toward Sgr\,A$^*$.  
The two red vertical lines at \mbox{v$_{\rm LSR}$\,$\simeq$$+$120 and
$+$270\,km$^{-1}$} delineate the HPVW emission.
The intensity scale is in $T_{\rm mb}$ units (baseline subtracted) except for \mbox{H$_2$O~1$_{1,0}$-1$_{0,1}$} that is in 
$T_{\rm mb}$/$T_{\rm C}$ (divided by the continuum).}
\label{fig:multi-line}
\end{figure}

 H$_2$O and C$^+$ also show HPVW  components \mbox{(Figure~\ref{fig:multi-line})}.
Owing to its 11.3\,eV ionization potential, the detection of [\CII]158\,$\mu$m  implies  that the high-velocity gas is irradiated by stellar far-ultraviolet photons 
(\mbox{FUV; $5\lesssim E < 13.6$~eV)}. The presence of H$_2$O~1$_{1,0}$-1$_{0,1}$ HPVW emission, and of several rotationally excited H$_2$O lines \citep{Goicoechea13} suggests elevated  temperatures, such as those in shocks; the gas-phase
routes leading to  abundant water vapor are endothermic or possess
energy barriers \citep[][]{vD13}.
Interestingly, the atomic carbon fine-structure lines \mbox{[\CI]492,\,809~GHz} do not show  HPVW. Only the ground-state line
at $\sim$492\,GHz displays an independent feature at v$_{\rm LSR}$\,$\simeq$$+$165\,km\,s$^{-1}$; the [\CI]809\,GHz line is barely detected in this velocity range. This feature 
is traditionally associated with the back side  of the
EMR.  Indeed, the very low \mbox{[\CI]809/492} line intensity ratio  $\lesssim$1 
 of this  feature (with $I_{\rm 492}$ and $I_{\rm 809}$ in W\,m$^{-2}$\,sr$^{-1}$) implies cold gas temperatures and thus is not related
to the source of HPVW emission. 

The HIFI scan also shows  \mbox{mid-$J$} HCO$^+$ and HCN emission lines (up to $J$=8-7). These lines do not display HPVW emission component either  (Fig.~\ref{fig:HCN}). Their lower velocity
emission range ($|$v$_{\rm LSR}$$|$$<$150\,km\,s$^{-1}$) and  high 
HCN/HCO$^+$ intensity ratios \citep[1.6-1.7; likely enhanced 
by IR pumping of vibrationally excited HCN lines, see e.g.,][]{Mills13} agree with  previous maps
of the CND in the same species \citep{Montero09,Mills13}. Given the large  size
of HIFI beams at these frequencies, very likely the HCN and HCO$^+$ emission  toward Sgr~A$^*$ is   beam-peaked emission
from the CND.

\section{Physical conditions of high-velocity CO gas}

\begin{figure}[t]
\centering
\includegraphics[scale=0.39, angle=0]{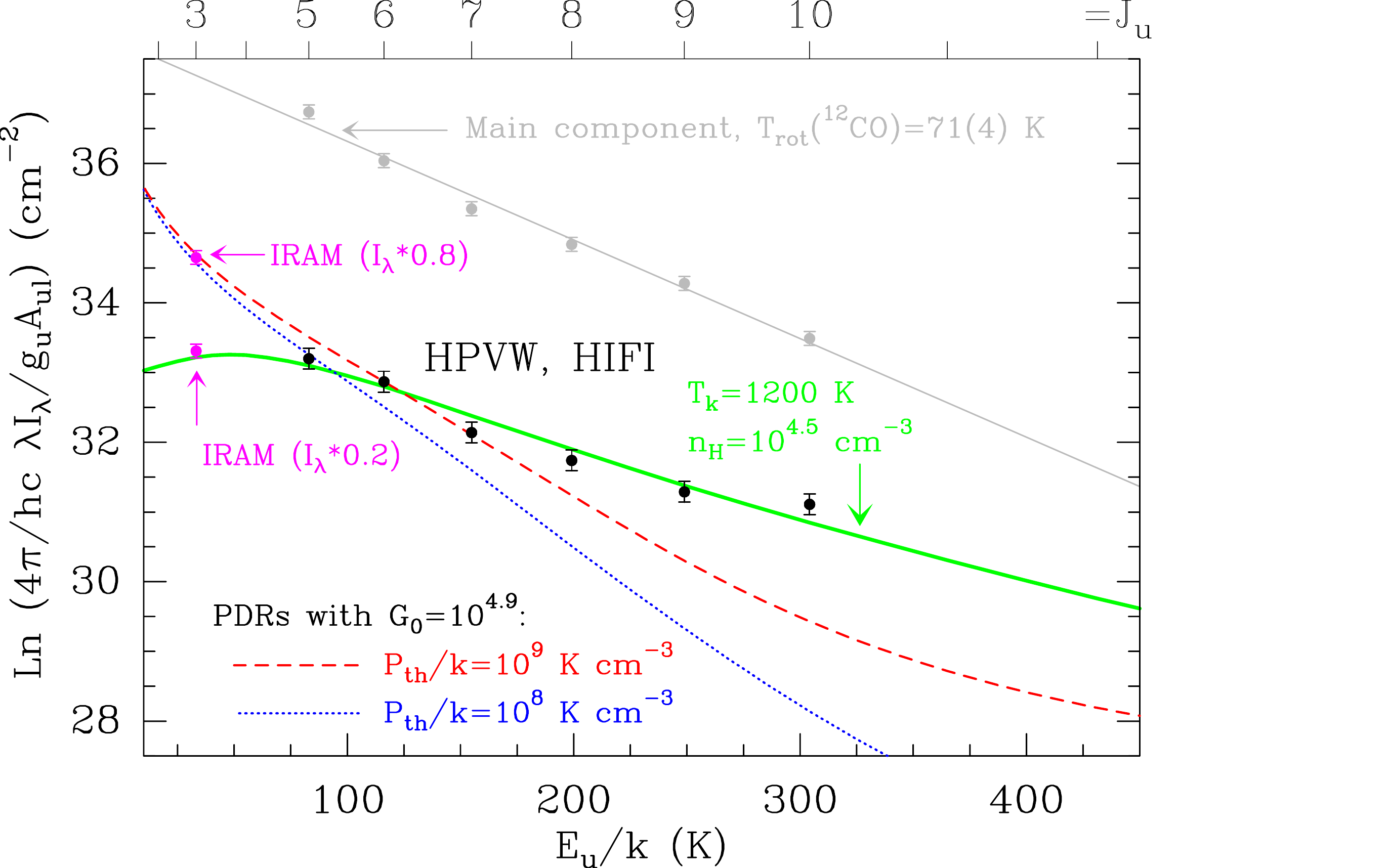} 
\caption{Observed CO intensities of the main and HPVW components toward
Sgr\,A$^*$ translated to rotational population diagrams. The green curve shows the best isothermal non-LTE model. The red and blue dashed curves show predictions of 
 PDR models with $G_0$=10$^{4.9}$ and
\mbox{$\zeta_{\rm CR}$=10$^{-15}$\,s$^{-1}$} (line intensities 
multiplied by 0.04).}
\label{fig:PDR_mods}
\end{figure}

From the detection of a few low-energy rotational lines (e.g.,~with ALMA) it is not easy to  constrain the physical conditions of the high-velocity cloudlets seen inside the cavity. With \mbox{\textit{Herschel}/HIFI} we spectrally resolve the HPVW emission in several rotationally excited CO lines.
From the observed lines, we  determine the beam-averaged $^{12}$CO 
column density ($N$) and rotational temperature ($T_{\rm rot}$) 
associated with the HPVW component, which has line intensities integrated from
$+$120 to $+$270~km\,s$^{-1}$(see \mbox{Table~\ref{table_wingsCO})}. 
\mbox{Table~\ref{table:rot-diag}} summarizes the inferred $N$($^{12}$CO) and 
\mbox{$T_{\rm rot}$(mid-$J$ $^{12}$CO)} values based on \mbox{rotational  diagram}
fits 
(\mbox{Figure~\ref{fig:PDR_mods}}). 
The HPVW  component  shows a higher degree of excitation 
(\mbox{$T_{\rm rot}$$\simeq$100~K versus 70~K}) and more pronounced curvature 
than the main spectral component,  i.e., total integrated line intensity minus HPVW. 
As for HCN and HCO$^+$, signal from the mid-$J$ $^{12}$CO main component is likely 
dominated by emission from the CND and surroundings (Fig.~\ref{fig:CND-image}).

Assuming a $^{12}$CO abundance  of
2$\cdot$10$^{-5}$ with respect to H nuclei \citep{Goicoechea18}, the $^{12}$CO column density  inferred from the HPVW \mbox{rotational diagram} is equivalent to 
$A_V$$\simeq$0.5~mag of visual extinction, or 2.5\,mag if we assume a uniform beam-filling factor of 0.2 for the cloudlet ensemble. These values agree with the average extinction estimated from  ALMA images \cite[][]{Goicoechea18} and
translate to a hot molecular gas mass of $\sim$10-60\,$M_{\odot}$ in the inner 1.5~pc of the Galaxy. This mass is  comparable to that  
 of neutral atomic \citep[$\sim$300\,M$_{\odot}$,][]{Jackson93} and  ionized gas
\citep[$\sim$60\,M$_{\odot}$, e.g.,][]{Lo83} inside the central  cavity.

Given the strong stellar FUV radiation field in the region \citep[$G_0$$\simeq$10$^{5.3}$ to 10$^{4.3}$ in Habing units,][]{Goicoechea18}, we might expect a dominant role of radiative heating. However, state-of-the-art
photodissociation region (PDR) models run with the Meudon code \citep{Petit06}
for $G_0$$\simeq$10$^{4.9}$, a cosmic-ray ionization rate of
\mbox{$\zeta_{\rm CR}$=10$^{-15}$\,s$^{-1}$}, and constant thermal pressures
\mbox{$P_{\rm th}/k$=10$^8$-10$^9$\,K\,cm$^{-3}$} do not  reproduce the shape of
the HPVW CO rotational ladder, especially \mbox{as $J$} increases (see Fig.~\ref{fig:PDR_mods}). Similar conclusions were anticipated by 
\citet{Goicoechea13} from (velocity unresolved) observations of  \mbox{higher $J$ CO} lines (up to \mbox{$J$=24-23}). These authors suggested that only a very
small filling factor ($<$0.05) ensemble of \mbox{FUV-irradiated} dense ($n_{\rm H}$$\approx$10$^7$~cm$^{-3}$)
clumps would be compatible with a dominant role of radiative heating (PDRs).  However, the filling factor of
the HPVW cloudlets determined from ALMA  is  larger and gas densities seem much lower; otherwise the cloudlets  would be  prominent in IR extinction maps. 

In order to constrain the range of average physical conditions that
reproduce the observed mid-$J$ CO lines \mbox{(HPVW component)}, we have run a grid of \mbox{non-local}, \mbox{non-LTE} excitation models (see Appendix~\ref{mtc_models}).
While only for very high densities ($n_{\rm H}$$>$10$^7$\,cm$^{-3}$) the gas is close to thermalization (\mbox{$T_{\rm rot}$$\simeq$$T_{\rm k}$$\simeq$100~K}), at lower densities   the excitation becomes \mbox{subthermal}  (\mbox{$T_{\rm rot}$=100~K$\ll$$T_{\rm k}$}), the implied gas temperature is higher, and the rotational population diagram shows curvature \citep{Neufeld12}. Our best fit is for 
$T_{\rm k}$$\simeq$1200~K and $n_{\rm H}$$\simeq$10$^{4.5}$~cm$^{-3}$ (green
curve in Fig.~\ref{fig:PDR_mods}). This model also fits the single-dish CO~$J$=3-2 emission fraction expected from the spatially unresolved high-velocity cloudlets (filling factor of $\simeq$0.2). We note, however, that a more accurate solution will require us to observe and spectrally resolve higher $J$ CO lines.

\begin{table}[t]
\caption{Beam-averaged rotational temperatures, column densities, 
and \mbox{correlation} coefficients obtained from rotational diagram fits.}   
\centering
\begin{tabular}{lccc}
\hline\hline
 Emission component                         & $T_{\rm rot}$  & $N$($^{12}$CO)     & r$^2$  \\
                                 &  (K)           & (cm$^{-2}$) & \\ \hline             
$^{12}$CO HPVW$^a$ & 101(12)      & 1.9(4)$\times 10^{16}$\,$^b$ & 0.946\\
$^{12}$CO main component      & 71(4)          & 6(1)$\times 10^{17}$ & 0.986 \\ 
\hline
\end{tabular} 
\tablefoot{$^a$From \mbox{v$_{\rm LSR}$=$+$120 to $+$270~km\,s$^{-1}$}.
\mbox{$^b$$N$($^{12}$CO)=9.6$\times 10^{16}$~cm$^{-2}$} if one assumes a uniform filling factor of 0.2.
}
\label{table:rot-diag} 
\end{table}

\citet{Goicoechea18} estimated  (assuming \mbox{$T_{\rm rot}$(CO 3-2)=100~K} precisely) density peaks around \mbox{$n_{\rm H}$$\simeq$(0.2-1.0)$\cdot$10$^5$~cm$^{-3}$} for  the  cloudlets resolved by ALMA. For these densities,
our non-LTE excitation models shown in \mbox{Figure~\ref{fig:MTC_mods}} imply  temperatures ranging from \mbox{$T_{\rm k}$$\simeq$400 to 2000~K}. 
 Both the elevated $T_{\rm k}$ and  low $n_{\rm H}$ 
(compared to the local Roche density) are compatible with
the  non-thermal H$_2$ level populations inferred  from 
IR observations with the \textit{VLT} \citep{Ciurlo16}.
These $T_{\rm k}$ and $n_{\rm H}$ also agree with  H$_2$ multi-line detections  \citep[$T_{\rm ex}$(H$_2$)$\sim$1100~K,][]{Mills17} with \textit{ISO}. Infrared observations provide lower
spectral resolution, thus it is difficult to extract the different velocity components independently. 
In summary, our observations add more quantitative evidence to the existence of high-velocity, hot  molecular cloudlets  inside the cavity of the CND and close to Sgr~A$^*$ ($<$1~pc). Although the detection of [\CII]158\,$\mu$m wing emission  demonstrates the presence of FUV illumination, PDRs alone cannot drive the
heating of the HPVW   component. 
Instead, \mbox{FUV-irradiated} shocks, which have $G_0$ values that are much higher than in the models of  \citet{Lesaffre13} and \citet{Melnick15},
likely dominate the heating of the hot molecular gas.

\begin{acknowledgements}

We are indebted to Tom Bell for initiating this project and for his help in the
early data reduction. We  thank our anonymous referee for useful comments. We also thank the Spanish MCIU and the ERC for funding support under grants AYA2017-85111-P and ERC-2013-Syg-610256-NANOCOSMOS.

\end{acknowledgements}


\bibliographystyle{aa}
\bibliography{references}

\begin{appendix}

\section{Non-local, non-LTE excitation isothermal models}\label{mtc_models}

To constrain the range of plausible physical conditions that reproduce
the mid-$J$ CO line excitation and  inferred $T_{\rm rot,\,HPVW}$ values, we run a grid of non-local and non-LTE
 models \citep[\mbox{Monte Carlo} model in Appendix of][]{Goicoechea_2006}. 
We included  \mbox{$^{12}$CO-H$_2$}  and  \mbox{$^{12}$CO-H} rotationally  inelastic collisions
\citep{Yang10,Balakrishnan02} and assumed a molecular gas fraction of 90\,\%, 
$n$(H)$=$0.25\,$n$(H$_2$); this value is consistent with a strongly \mbox{FUV-irradiated} gas
in which not all hydrogen is in molecular form.
We adopted a non-thermal velocity dispersion of
\mbox{$\Delta$v=35~km\,s$^{-1}$} (from turbulence and macroscopic gas motions). This
value is the observed average linewidth of the cloudlets detected
by ALMA. 
We take the same $N$($^{12}$CO) obtained from the rotational diagram 
 and search for the range of $T_{\rm k}$ and
$n_{\rm H}$=$n$(H)+2$n$(H$_2$) values that reproduce the derived  
\mbox{$T_{\rm rot,\,HPVW}$$\simeq$100~K} and observed
line intensities.

\begin{figure}[h]
\centering
\includegraphics[scale=0.48, angle=0]{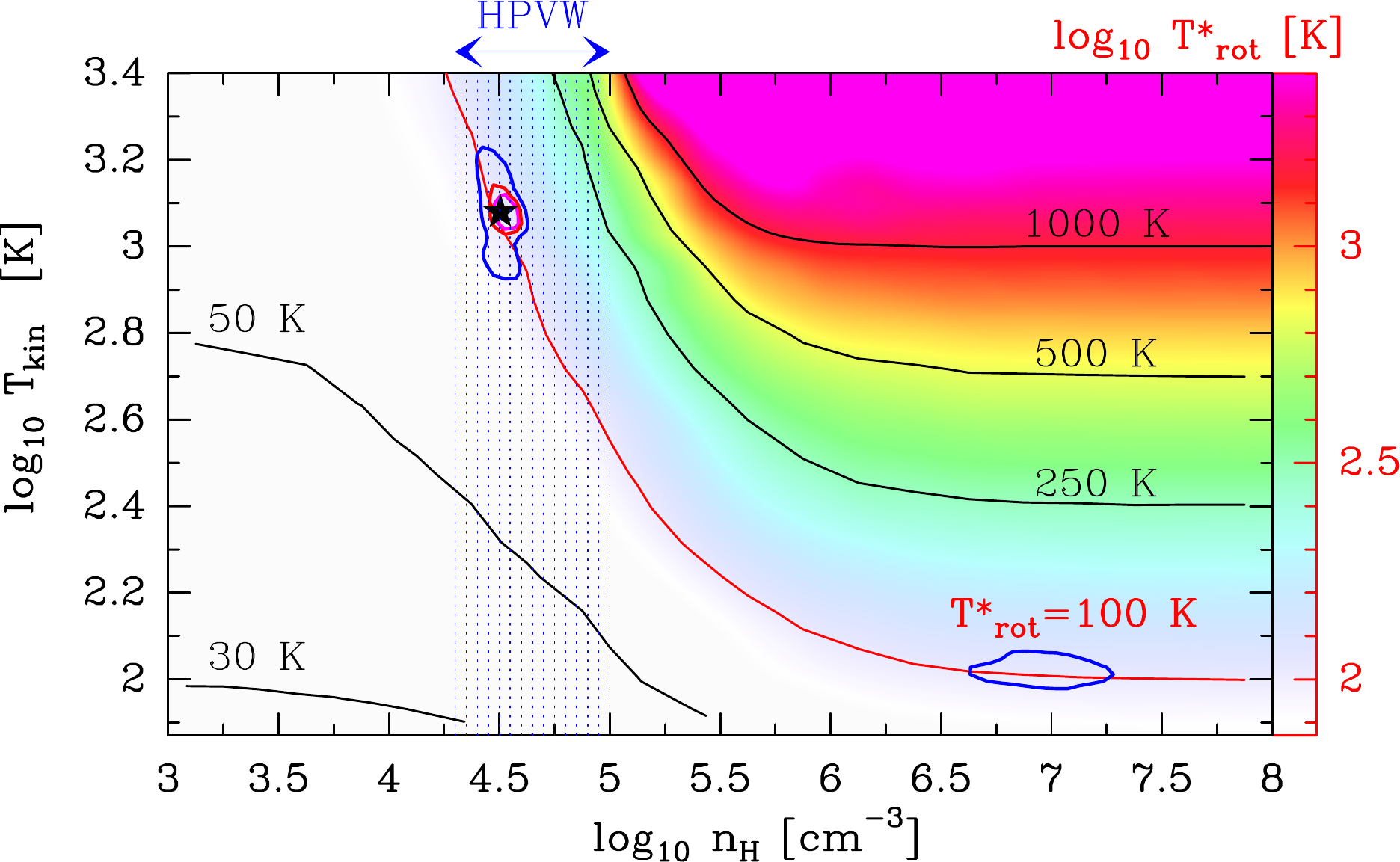} 
\caption{$^{12}$CO rotational temperatures ($T^{*}_{\rm rot}$, computed by fitting a straight line to synthetic line intensities in the \mbox{$J_{\rm up}$=5-10}  range)
obtained from  isothermal  non-LTE models. The red curve shows the parameter
space that reproduces the observed $T_{\rm rot}$ of the HPVW component. The blue-shaded
area shows the estimated gas density of the high-velocity  cloudlets detected by ALMA \citep{Goicoechea18}. The magenta, red, and blue curves show
rms$_{\rm min}$/rms contours of 1, 0.95, and 0.9 respectively, where 
rms refers to the rms value of \mbox{log$_{10}$\,(I$_{\rm HIFI}$/I$_{\rm mod}$}).
}
\label{fig:MTC_mods}
\end{figure}

Figure~\ref{fig:MTC_mods} shows isothermal model results in the form of iso-$T^{*}_{\rm rot}$ contours. For each model we created rotational diagrams and determined
a rotational temperature ($T^{*}_{\rm rot}$) by fitting a 
\mbox{straight line}\footnote{At low densities, \mbox{$n_{\rm H} \ll n_{\rm cr}$($J$)}, CO  rotational level populations are such that plotted in a rotational diagram they show considerable curvature  \citep[][]{Neufeld12}. We also fit these models with a straight line.}  to the synthetic CO line intensities in the $J_{\rm up}$=5–10 range.
To estimate the fit goodness we calculated the minimum rms value of 
\mbox{log$_{10}$(I$_{\rm HIFI}$/I$_{\rm mod}$)}.
\mbox{Figure~\ref{fig:MTC_mods} shows}  rms$_{\rm min}$/rms contours of 1, 0.95 and 0.9  
in magenta, red, and blue, respectively. The best model (rms=0.076, indicated with a star in the figure) implies \mbox{$T_{\rm k}$$\simeq$1200~K} and \mbox{$n_{\rm H}$$\simeq$10$^{4.5}$~cm$^{-3}$}, but we note that a  more accurate solution would require velocity-resolved observations of \mbox{higher $J$ CO} lines. 

Given the intense IR, visible, and UV radiation fields in the cavity, one may consider
whether radiative pumping to CO vibrational and electronic levels  can influence the emergent
rotationally excited line intensities.
For moderate gas densities ($>$10$^4$\,cm$^{-3}$) inelastic collisions are important drivers of CO rotational excitation. Thus very strong  mid-IR fields
are needed to alter the rotational level populations \citep[e.g.,][]{Carroll81,Godard13}. 
The estimated mid-IR color temperatures in the cavity range from $\approx$200~K 
\citep{Cotera99} to \mbox{145-85~K} \citep{Lau13}.
These values are around the 160~K threshold estimated by \citet{Carroll81}
for CO pumping  in the no-collision limit. Hence, we suspect that IR pumping can contribute 
locally, around evolved stars or in protostellar envelopes, but does not drive the excitation of the optically thin mid-$J$ CO lines we detect at high velocities.
Indeed,  IR observations toward the cavity do not report high positive-velocity features
in the wings of the saturated \mbox{$^{12}$CO $v$=1-0} ro-vibrational lines
\citep{Goto14}.  

Similarly, \mbox{visible/UV} pumping  may have
some relevance in strongly irradiated environments. \mbox{Visible/UV} pumping
(included in the PDR models shown in Figure~\ref{fig:PDR_mods})
can enhance the intensity of the CO rotationally excited lines only by a factor of a few \citep{Godard13}. 
Either way, these radiative excitation processes  further  \textit{decrease} the  density of the HPVW gas component below our estimation from the grid of non-LTE excitation  models. Therefore, the high gas density solutions, \mbox{$n_{\rm H}$$\simeq$10$^7$~cm$^{-3}$}
(shown in Figure~\ref{fig:MTC_mods}), do not seem favored
for the molecular gas in the cavity.


\section{Complementary figures and tables}

\begin{table*}[h]
\caption{\label{table:detected_line}Spectroscopic and observational parameters of 
the lines discussed in this work.} 
\centering
\resizebox{\textwidth}{!}{%
\begin{tabular}{lcrrclcccr}
\hline\hline
        &                              &  Frequency  & $E_{\rm u}/k$ & $A_{\rm ul}$ & Intensity$^a$, $I$  & $<$v$>$$^b$ & $<$$\Delta$v$>$$^c$   &Beam &  \textit{Herschel} \\
Species & Transition                   &   \multicolumn{1}{c}{(GHz)}    &   \multicolumn{1}{c}{(K)}        & (s$^{-1}$)   & (W\,m$^{-2}$\,sr$^{-1}$)  & (km\,s$^{-1}$) & (km\,s$^{-1}$) &  (arcsec) & ObsID  \\
\hline
C$^+$    &    $^{2}P_{3/2}$ - $^{2}P_{1/2}$ &  1900.53690  & 91.2  & 2.3 $\times 10^{-6}$  & 8.56 $\times 10^{-6}$ & 22.0 & 155.3 &   11  & 1342243685 \\   
C        & $^{3}P_{2}$ - $^{3}P_{1}$     &  809.90231      & 62.5  & 2.7 $\times 10^{-7}$  & 1.83 $\times 10^{-7}$ & 18.4 & 135.1 & 26 & 1342230279 \\
         & $^{3}P_{1}$ - $^{3}P_{0}$       &  492.50139 & 23.6 & 7.9 $\times 10^{-8}$  & 4.73 $\times 10^{-8}$ & 24.0  & 155.3 & 42 & 1342249609 \\
CO       & $J$=10-9                    & 1151.98544   &  304.2  &   1.0 $\times 10^{-4}$  & 5.00 $\times 10^{-7}$ & 32.0 & 132.2 & 19 & 1342251185\\ 
         & $J$=9-8                     & 1036.91239   &  248.9  &   7.5 $\times 10^{-5}$  & 6.27 $\times 10^{-7}$ & 17.1 & 123.2 & 20 & 1342243700\\
         & $J$=8-7                     &  921.79970   &  199.1  &   5.2 $\times 10^{-5}$  & 6.05 $\times 10^{-7}$ & 18.3 & 148.6 & 23 & 1342243707\\
         & $J$=7-6                     &  806.65180   &  154.9  &   3.5 $\times 10^{-5}$  & 5.14 $\times 10^{-7}$ & 31.3 & 118.2 & 26  & 1342230279 \\
         & $J$=6-5                     &  691.47208   &  116.2  &   2.2 $\times 10^{-5}$  & 4.79 $\times 10^{-7}$ & 21.6 & 131.7 & 30 & 1342239594  \\
         & $J$=5-4                     &  576.26793   &   83.0  &   1.2 $\times 10^{-5}$  & 3.83 $\times 10^{-7}$ & 23.4 & 129.1 & 36 & 1342230394  \\ 
$^{13}$CO& $J$=6-5                     &  661.06728   &  111.1  &   1.9 $\times 10^{-5}$  & 3.08 $\times 10^{-8}$ & 13.6 & 143.0 & 32 & 1342239594 \\
         & $J$=5-4                     &  550.92628   &   79.3  &   1.1 $\times 10^{-5}$  & 3.28 $\times 10^{-8}$ & 14.6 & 161.4 & 38 & 1342249609 \\
C$^{18}$O& $J$=5-4                     &  548.83101   &   79.0  &   1.1 $\times 10^{-5}$  & 2.41 $\times 10^{-9}$ & 71.9 & 92.8  & 38 & 1342249609\\ 
CN     & $N$,$J$= 5,11/2-4,9/2 & 566.94721 & 81.6   &   1.9 $\times 10^{-3}$  & 5.50 $\times 10^{-9}$ & 44.7 & 130.3 & 37 & 1342249609 \\
HCN      & $J$=8-7                     &  708.87700   &  153.1  &   1.7 $\times 10^{-2}$  & 8.74 $\times 10^{-9}$ & 37.6 & 110.4 & 30 & 1342239594\\
HCN      & $J$=7-6                     &  620.30400   &  119.1  &   1.2 $\times 10^{-2}$  & 4.45 $\times 10^{-9}$ & 31.3 & 90.6   & 34 & 1342230394\\
HCN      & $J$=6-5                     &  531.71635   &  89.3   &   7.2 $\times 10^{-3}$  & 5.82 $\times 10^{-9}$ & 28.8 & 92.4  & 39 & 1342249609 \\
HCO$^+$  & $J$=7-6                     &  624.20836   &  119.8  &   2.0 $\times 10^{-2}$  & 4.45 $\times 10^{-9}$ & 30.2 & 112.0  & 33 & 1342230394 \\
HCO$^+$  & $J$=6-5                     &  535.06158   &  89.9   &   1.2 $\times 10^{-2}$  & 5.82 $\times 10^{-9}$  & 30.8 & 119.4 & 39 & 1342249609  \\
$o$-H$_2$O & $J_{Kk}$=1$_{10}$-1$_{01}$&  556.93600   &  26.7   &   3.5 $\times 10^{-3}$  & 9.28 $\times 10^{-9}$  & 68.8 & 149.0 & 37 & 1342249609 \\
\hline
\end{tabular}
}
\tablefoot{$^a$Moment~0 or total line intensity over each emission profile 
\mbox{$I=$$\sum I({\rm v})\Delta $v}. Total uncertainty up to $\sim$10$\%$.
 $^b$Moment~1: Intensity-weighted mean velocity \mbox{$<$v$>$=$\sum I({\rm v})\,{\rm v}/
\sum$v}.$^c$Moment~2: Intensity-weighted velocity dispersion 
\mbox{$<$$\Delta$v$>$=$\sqrt{\frac{\sum [I({\rm v})({\rm v}-{\rm M1})^2]}{\sum I({\rm v})}}$}.} 
\end{table*}

\begin{table}[h]
\caption{\label{table_wingsCO} Line intensities for the HPVW  component.} 
\centering
\begin{tabular}{lccccc}
\hline\hline
         &               &   Intensity$^a$, $I$  \\
Species  & Transition    &  (W\,m$^{-2}$\,sr$^{-1}$) \\
\hline
CO         & $J$=10-9                    & 4.26 $\cdot 10^{-8}$\\
           & $J$=9-8                     & 3.00 $\cdot 10^{-8}$\\
           & $J$=8-7                     & 2.62 $\cdot 10^{-8}$\\
           & $J$=7-6                     & 2.00 $\cdot 10^{-8}$\\
           & $J$=6-5                     & 1.92 $\cdot 10^{-8}$\\
           & $J$=5-4                     & 1.08 $\cdot 10^{-8}$\\
C$^+$      & $^{2}P_{3/2}$-$^{2}P_{1/2}$ & 1.49 $\cdot 10^{-8}$   \\
$o$-H$_2$O & $J_{Kk}$=1$_{10}$-1$_{01}$  & 1.86 $\cdot 10^{-9}$   \\
\hline
\end{tabular}
\tablefoot{$^a$Integrated from v$_{\rm LSR}$=$+$120 to 270 km s$^{-1}$.\\
Total uncertainty up to $\sim$15$\%$.}
\end{table}

\begin{figure}[b]
\centering
\includegraphics[scale=0.6, angle=0]{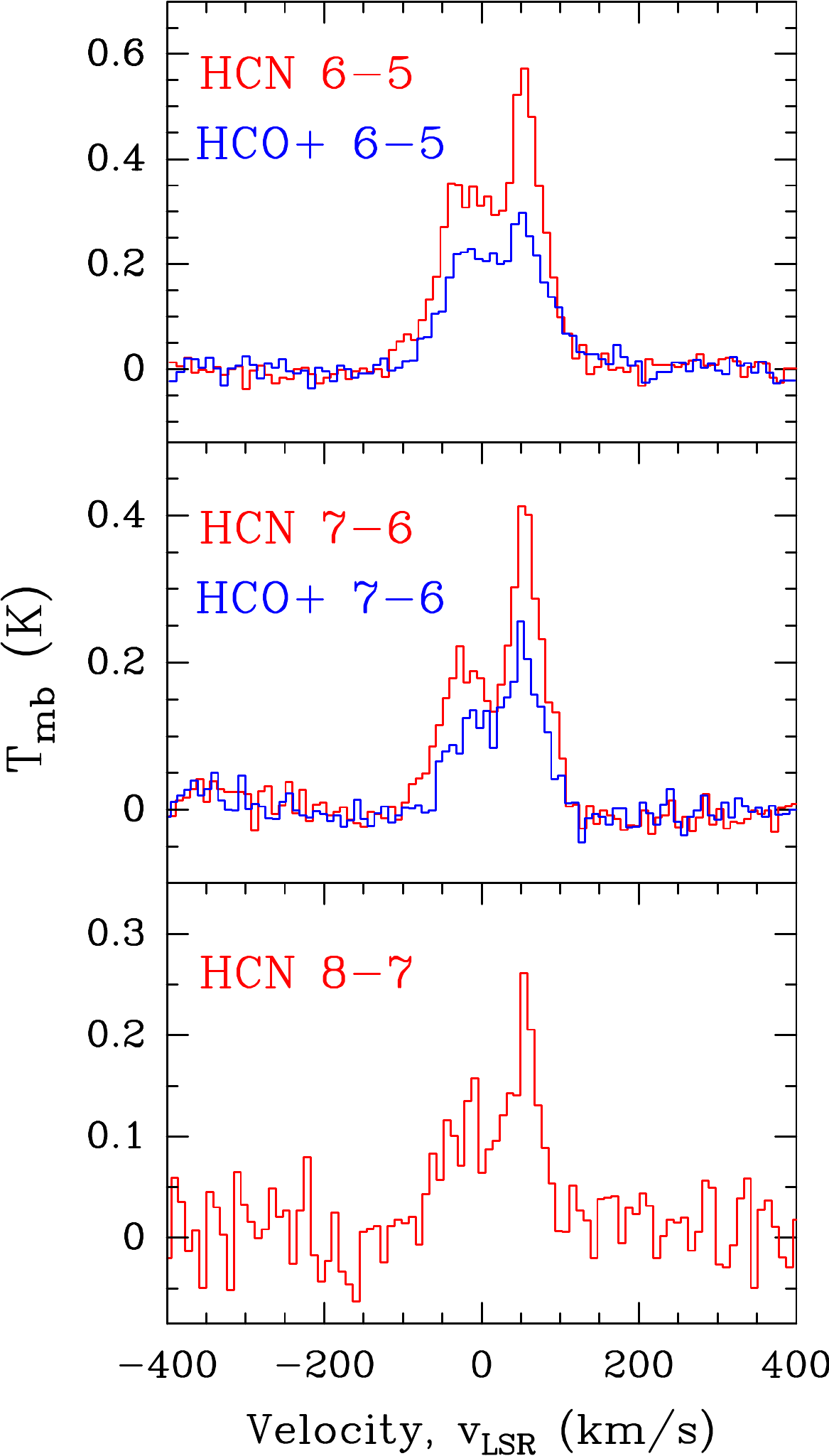} 
\caption{HIFI detection of HCN and HCO$^+$ lines  toward Sgr~A$^*$
showing emission only at velocities typical of the CND
(\mbox{v$_{\rm LSR}$$<$150\,km\,s$^{-1}$}).}
\label{fig:HCN}
\end{figure}

\end{appendix}

\end{document}